\documentclass[aps,epsf,amsmath,superscriptaddress,citeautoscript,twocolumn,
showpacs,floatfix]{revtex4-2}
\usepackage{bm}
\usepackage{amssymb, float}
\usepackage{graphicx}
\usepackage{amsmath, amsthm, amssymb, graphicx}
\usepackage[usenames]{color}
\graphicspath{{figures/}} 
\usepackage{ulem}

\usepackage{amsmath,amsthm,amsfonts,amscd,amssymb}
\usepackage{eucal} 	 	
\usepackage{verbatim}      	
\usepackage{makeidx}       	
\usepackage{braket}

\usepackage{natbib}
\usepackage{bibentry}
\usepackage{hyperref}

\usepackage{mathrsfs}
\usepackage{graphicx}
\usepackage{dcolumn}
\usepackage{bm}

\begin{document}
	
	\preprint{}
	
	\title{Two Toy Spin Chain Models of Decoherence}

	\author{ P.C.E. Stamp}
	\affiliation{Department of Physics and Astronomy, University
		of British Columbia, 6224 Agricultural Rd., Vancouver,
B.C., Canada
		V6T 1Z1}
	\affiliation{Pacific
		Institute of Theoretical Physics, University of British
Columbia,
		6224 Agricultural Rd., Vancouver, B.C., Canada V6T 1Z1}
	
\author{Zhen Zhu}	
	\affiliation{Pacific
		Institute of Theoretical Physics, University of British
Columbia,
		6224 Agricultural Rd., Vancouver, B.C., Canada V6T 1Z1}
	
\vspace{5mm}

	\begin{abstract}

We solve for the decoherence dynamics of two models in which a
simple qubit or `Central Spin' couples to a bath of spins; the bath
is made from a chain of spins. In model 1, the bath spins are Ising
spins; in Model 2, they are coupled by transverse spin-spin
interactions, and the chain supports spin waves. We look at (i) the
case where the Hamiltonian is static, with a constant system/bath
coupling, and (ii) where this coupling varies in time.

	\end{abstract}
	
	\maketitle


\section{Introduction}


The dynamics of `decoherence', ie., the gradual entanglement of a
quantum system with its environment - is of considerable subtlety. In
the simplest picture, one imagines some sort of exponential decay in
time of off-diagonal matrix elements in a reduced density matrix for
the central system. For a single central spin or `qubit', this would
occur on a timescale $T_2$, distinct from the time $T_1$ for the decay
of the diagonal matrix elements.

This simple picture is only correct under very restrictive conditions
- it is correct, eg., for a central spin coupled to random noise.
Solutions for the dynamics of the `spin-boson' model
\cite{AJL87,weiss} (where the central qubit couples to an oscillator
bath) or the `central spin' model \cite{PS00} (where the central spin
couples to a bath of spins) show much more complicated behaviour. In
the central spin model, one can  see quite complex non-monotonic
behaviour in time of the density matrix elements. If one then looks at
the correlations between the central system and the environmental
variables, extremely complex features emerge, in which entanglement
correlators at different levels exchange information in interesting
ways \cite{cox18,coxThesis}.

Even more complex features can emerge when the central system goes
beyond a simple two-level `qubit' system, to some object hopping on
some lattice. This is a topic of considerable interest to those
interested in the decoherence dynamics of quantum information
processing systems (a huge topic), in the dynamics of quantum walks
\cite{venegas12,PS06,carlstrom16}, in the well known problem of motion
of holes in insulating magnetic systems
\cite{carlstrom16,nagaoka,brinkman,greiner21,neilsen22}, and anyone
interested in general problems in quantum diffusion.

\vspace{2mm}

Quite generally the models that are employed fall into 2 classes,
viz.,

\vspace{2mm}

(i) {\it Oscillator Bath Models}: In such models the environment is
modelled as a set of independent oscillators \cite{weiss}; these are
supposed to represent extended modes (like phonons, photons,
electron-hole pairs, gravitons, etc.). For a system containing $N$
such oscillators, the coupling of the system to each oscillator is
$\propto N^{-1/2}$, so that for large $N$ (or for $N \rightarrow
\infty$) the effect of the bath on the central system is independent
of $N$, and one can assume an equilibrium bath at some temperature
$T$.

\vspace{2mm}

(ii) {\it Spin Bath Models}: The environment is now modelled as a set
of `pseudospin' degrees of freedom \cite{PS00}, each possessing a
finite-dimensional Hilbert space (often a 2-dimensional space, so that
the environment is a set of 2-level systems). These represent
localized modes (like defects, nuclear or paramagnetic spins, or
dynamic impurities in the system). In this case the coupling between
the central system and each bath `spin' may not depend on $N$ at all,
and certainly there is no requirement that it be $\propto N^{-1/2}$.
One cannot assume that this bath is in equilibrium.

\vspace{2mm}


\begin{figure}[H]
\includegraphics[width=9cm]{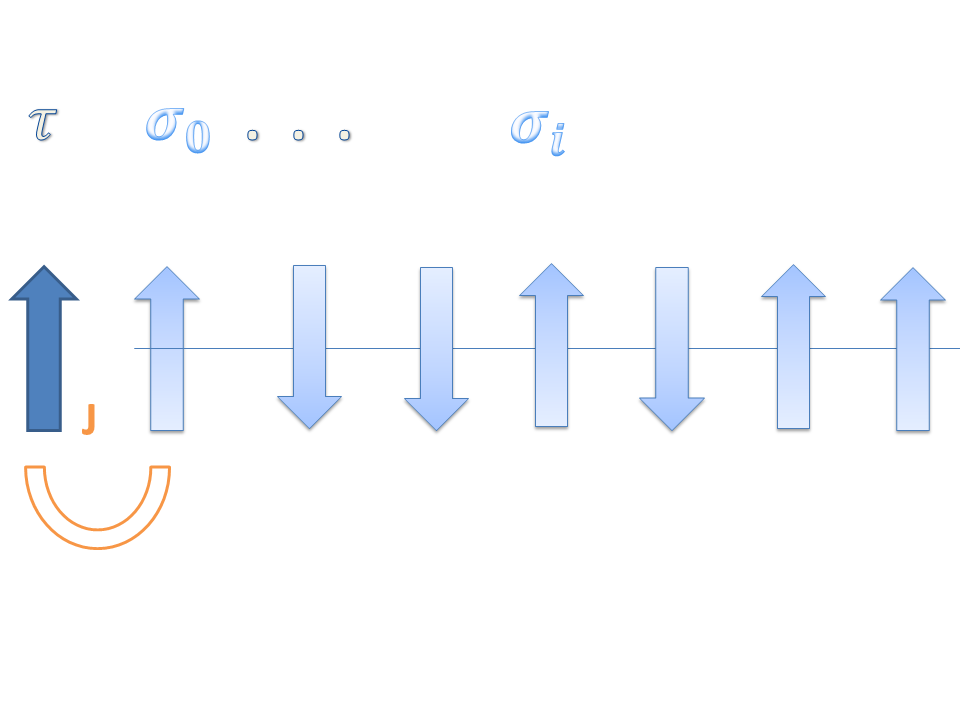}
\caption{\label{fig:path_5} Illustration of our spin chain model. A
central spin  couples to a spin chain.}
\end{figure}


For an elementary discussion of the difference between these two
baths, see, eg., ref. \cite{gaita,SHPMP}. A key difference is that for
oscillator baths, decoherence is closely tied to dissipation: one
expects a fluctuation/dissipation connection to exist. However for
spin baths, no such necessary connection exists - in fact, one can
have very large decoherence with almost no dissipation. This makes
spin baths much more dangerous for, eg., quantum computation.

In this paper we wish to show some very basic features of decoherence
dynamics, by choosing two simple  `toy' models. The models are of some
pedagogical interest, just because they are so simple - indeed, they
are sufficiently simple that in one case we easily find an exact
solution, and in the other case a very accurate solution in the regime
of interest. In this short contribution we first introduce the models,
and give the details of their solutions, and then make some
observations based on these solutions. It will be seen that even in
these simple models (about as simple as one can get) there are several
interesting features, which persist in more complicated models.

It will be clear that much more can be said, which we have no space
for in this short communication. We will also have no space to discuss
the application of models of decoherence that are similar to the ones
discussed here. For recent discussions of both anomalous diffusion and
localization, which can occur when the decoherence is non-Markovian,
work on optical lattices is of some interest
\cite{preiss15,cooper18,major18,afek23}; models rather close in spirit
to the models discussed here are examined in ref.
\cite{PS06,carlstrom16,danaci21}. More general work on entanglement
and decoherence in condensed matter systems is reviewed by LaFlorencie
\cite{laflor16}.

In both models, the environment is modelled as a 1-D spin chain
$\{\sigma_i\}$, and the central system is a single spin $\tau$, which
is coupled to the first spin in the chain. The models differ only in
the form of the bath Hamiltonian.


\section{Model 1}

The first model has an environment or bath with `Ising' Hamiltonian,
coupled transversely to a qubit, via a single coupling $J$ between the
qubit and the first spin on the spin chain. The Hamiltonian is:
\begin{equation}\label{eq:H_1}
H=J \tau^y\sigma_0^y+V\sum_{i=0}^{N-1}\sigma_i^z\sigma_{i+1}^z,
\end{equation}
where $\tau^y$ is the central spin operator and the $\{\sigma_i\}$ are
operators for $N$ bath spins; $J$ is the coupling between the central
qubit and the first bath spin, and $NV$ is the band width of the bath
spins. We do {\it not} use periodic boundary conditions here.

The central spin has the reduced density matrix
\begin{equation}
\begin{pmatrix}
|a|^2 & a^* b \kappa(t) \\
ab^* \kappa^*(t)& |b|^2
\end{pmatrix}.
\end{equation}

This model can be solved exactly. Since $H$ commutes with $\tau^y$,
the diagonal component (represented in $\tau^y$ eigenstates) of the
reduced density matrix does not change in time. We write the
off-diagonal components $\rho_{12}(t)$ and $\rho_{21}(t)$ as
\begin{equation}
\rho_{12}(t)=\kappa(t)\rho_{12}(0),
\end{equation}
with $\kappa(t)$ being the ``decoherence factor"
\begin{equation}
\kappa=\mbox{tr}_b\left(e^{-iH^{+} t} \rho_b e^{-iH^{-} t}\right),
\end{equation}
where $H^{\pm}$ are the block Hamiltonians in the $\tau^y=\pm 1$
subspaces respectively:
\begin{equation}
H^{\pm}\;=\; \pm J \sigma_0^y +
V\sum_{i=0}^{N-1}\sigma_i^z\sigma_{i+1}^z
\end{equation}

The solution to this model is
\begin{eqnarray}
 \label{eq:exact}
\kappa(t) &=&\frac{V^2+J^2 \cos 2\sqrt{1+J^2/V^2}t }{V^2+J^2}
\nonumber \\
&=& 1-\frac{2J^2 \sin^2 \sqrt{1+J^2/V^2}t}{V^2+J^2}.
\end{eqnarray}
which is clear enough - one is basically dealing with a 2-spin
problem, because there is no coupling between the spins on the chain.
Thus all correlations and information existing initialy in the qubit
are confined to the qubit/bath spin pair.

It is also interesting to obtain an approximate solution to this
problem. To do this we use a Jordan-Wigner transformation to
diagonalize $H$, by writing
\begin{eqnarray}
\sigma_i^z & = &-\prod_{j<i}(1-2c_j^{\dag}c_j)(c_i^{\dag}+c_i),\\
\sigma_i^x & = & 1-2c_i^{\dag}c_i,\\
\sigma_i^y & = & i\prod_{j<i}(1-2c_j^{\dag}c_j)(c_i^{\dag}-c_i).
\end{eqnarray}

Substituting this into \eqref{eq:H_1} gives
\begin{eqnarray}
H &=& iJ \tau^y(c_0^{\dag}-c_0) \nonumber \\
&& \qquad
+V\sum_{i=0}^{N-1}(c_i^{\dag}c_{i+1}^{\dag}-c_ic_{i+1}^{\dag}+c_i^{\dag}c_{i+1}-c_i
c_{i+1}).
\end{eqnarray}
Introducing the Fourier transform $c_{k_n}=\frac{1}{\sqrt{N}}\sum_j
c_j e^{ijk_n}$, with $k_n =\frac{2\pi n}{N}$, and the Bogoliubov
transformation
\begin{eqnarray}
\gamma_{k_n}&=&e^{-\frac{ik_n}{2}}\left(\cos\frac{k_n}{2}c_{k_n}-i\sin\frac{k_n}{2}c_{-k_n}^{\dag}\right),\\
c_{k_n}&=&\cos\frac{k_n}{2}e^{\frac{ik_n}{2}}\gamma_{k_n}+i\sin\frac{k_n}{2}e^{\frac{ik_n}{2}}\gamma_{-k_n}^{\dag}.
\end{eqnarray}
the Hamiltonian becomes
\begin{equation}\label{eq:H_1_trans}\begin{split}
H&=\sum_n
\left(V(2\gamma_{k_n}^{\dag}\gamma_{k_n}-1)+\frac{iJ}{\sqrt{N}}\tau^y(\gamma_{k_n}^{\dag}-\gamma_{k_n})\right)\\
&=\sum_n H_n,
\end{split}
\end{equation}
with $ H_n=
V(2\gamma_{k_n}^{\dag}\gamma_{k_n}-1)+\frac{iJ}{\sqrt{N}}\tau^y(\gamma_{k_n}^{\dag}-\gamma_{k_n})$.

Now the individual Hamiltonians $H_n$ do not commute with each other;
we have
\begin{equation}\label{eq:commutator}
[H_n,H_{n'}]=-\frac{2J^2}{N}(\gamma_{k_n}^{\dag}-\gamma_{k_n})(\gamma_{k_{n'}}^{\dag}-\gamma_{k_{n'}}).
\end{equation}
However, in the weak interaction limit $J/\sqrt{N}V\ll 1$, these
commutators can be neglected (see Appendix), and the Hamiltonian
\eqref{eq:H_1_trans} becomes block diagonal. In each
$\{\gamma_{k_n}^{\dag}, \gamma_{k_n}^{}\}$ subspace, we have
\begin{eqnarray}
H_n(\tau^y=+1)&=&\begin{pmatrix}
1 & \frac{iJ}{\sqrt{N}V}\\
-\frac{iJ}{\sqrt{N}V} & -1
\end{pmatrix}\\
H_n(\tau^y=-1)&=&\begin{pmatrix}
1 & -\frac{iJ}{\sqrt{N}V}\\
\frac{iJ}{\sqrt{N}V} & -1
\end{pmatrix}.
\end{eqnarray}

Let's assume the bath is initially in a fully mixed state
$\rho_B=\frac{1}{2}\mathbf{I}$.  It is then straightforward to
calculate $\kappa(t)$:
\begin{equation}
\begin{split}
\kappa &=1-2\sin^2(\sqrt{1+\frac{J^2}{NV^2}}t)\frac{J^2}{NV^2+J^2}\\
&\approx 1-2\sin^2 t \frac{J^2}{NV^2}.
\end{split}
\end{equation}
so that as $N\to\infty$, we have:
\begin{equation}\label{eq:kappa1}
\kappa= \lim_{N\to \infty}\left(1-2\sin^2 t
\frac{J^2}{NV^2}\right)^N=e^{-2J^2\sin^2 t/V^2}.
\end{equation}
which gives the correct result up to order  $J^2/V^2$.


\begin{figure}[H]
\includegraphics[width=9cm]{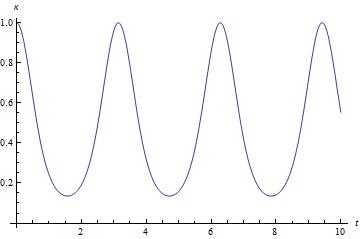}
\caption{\label{fig:kappa1}  Decoherence factor  $\kappa(t)$
\eqref{eq:kappa1} of model 1 as a function of time. There is no real
decoherence and $\kappa$ goes back to 1 periodically. }
\end{figure}


We see that $\kappa$ is periodic in time (see Fig. \ref{fig:kappa1}).
This is because, as already noted,  information cannot flow into the
whole spin chain - it is {\it confined} to the first local bath spin
and the central spin.

\section{Model 2}

We now consider another model in which the spin chain is the
environment, but in which there are interactions between the bath
spins (here, simple transverse spin-spin coupling). The model
Hamiltonian is
\begin{equation}\label{eq:H_2}
H=J
\tau^y\sigma_0^y+V\sum_{i=0}^{N-1}(\sigma_i^{+}\sigma_{i+1}^{-}+h.c.).
\end{equation}
in which spin waves can propagate down the chain. Thus the environment
is now a set of oscillators (spin waves); this model is a variant of
the spin-boson model.

\begin{figure}[H]
\includegraphics[width=9cm]{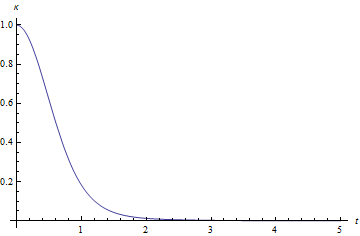}
\caption{\label{fig:kappa2} This is the decoherence factor of model 2
as a function of time. The loss of coherence is clear over time. }
\end{figure}

Again using the Jordan-Wigner transformation, we have
\begin{equation}
H=\sum_n \left(2V\cos
k_n(\gamma_{k_n}^{\dag}\gamma_{k_n}-\frac{1}{2})-\frac{iJ}{\sqrt{N}}\tau^y(\gamma_{k_n}^{\dag}-\gamma_{k_n})\right).
\end{equation}
which has the same structure as \eqref{eq:H_1_trans}, except now with
bath spectrum $\cos k_n V$ instead of a constant. We now follow the
same approximate procedure used for model 1, to get the decoherence
factor up to order $J^2$:
\begin{widetext}
\begin{equation}
\begin{split}
\frac{1}{2}\mbox{tr}\left(e^{-iH_n^{-}t}e^{-iH_n^{\dag}t}\right)
&=1-\frac{2\sin^2\sqrt{V^2\cos^2 k_n+\frac{J^2}{N}}t}{\cos^2
k_n+\frac{J^2}{NV^2}}\cdot\frac{J^2}{NV^2}\\
&\approx  1-\frac{2\sin^2(Vt\cos k_n )J^2}{N V^2\cos^2 k_n};\quad\quad
(N \to \infty)\\
&\approx  e^{-\frac{2\sin^2(Vt\cos k_n )J^2}{N V^2 \cos^2 k_n}}.
\end{split}
\end{equation}

After including all the $H_n$s, we finally get
\begin{equation}
\kappa (t) = \exp\left(-\sum_n \frac{2\sin^2(Vt\cos k_n )J^2}{NV^2
\cos^2 k_n}\right)
\end{equation}

If we now take the limit $N \rightarrow \infty$, this reduces to
\begin{equation}
 \label{eq:kappa2}
\begin{split}
\kappa (t) &
\approx\exp\left(-\frac{J^2}{\pi V^2}\int^{\pi}_{-\pi}dk
\frac{\sin^2(Vt\cos k )}{\cos^2 k}\right);\quad\quad  (N \to \infty)\\
&=\exp\left(-2
J^2t^2\,_1F_2\left(\frac{1}{2};\frac{3}{2},2;-V^2t^2\right)\right).
\end{split}
\end{equation}
where $_1F_2$ is a hypergeometric function.

The resulting behaviour of $\kappa(t)$ is illustrated in Fig.
\ref{fig:kappa2}. Its long time asymptotic behaviour can still be
studied. When $t\to\infty$, we can expand the generalized
hypergeometric function as
\begin{equation}
(Vt)^2\,_1F_2\left(\frac{1}{2};\frac{3}{2},2;-V^2t^2\right) \;\;=\;\;
Vt \;+\; \frac{1}{2}\sqrt{\frac{1}{\pi
Vt}}\cos\left(2Vt-\frac{\pi}{4}\right)+O\left(\frac{1}{(Vt)^{\frac{3}{2}}}\right).
\end{equation}
\end{widetext}
and we see that $\kappa\to e^{-J^2 t/V}\to 0$  as $t\to\infty$.

We see that in this model the system decoheres completely as $t
\rightarrow \infty$;  the quantum information is forever lost to the
environment, into the spin wave bath. In fact, we are dealing here
with a particular species of spin-boson model.


\section{Time-Varying System-Bath Coupling}


What if we are allowed to switch on or off the coupling between the
central system and the bath? One can envisage several such situations
- for example, we can simply switch the coupling off, or we can switch
it off and then bring it back again, in an effort to reverse the
decoherence.

\subsection{Switch-off process}

Let us assume a time-varying coupling $J(t)$ of form:
\begin{equation}
H=J(t)\tau^y\sigma_0^y+H_{bath},
\end{equation}
where $J(t) \to J_0$ as $t\to -\infty$ and $J(t) \to 0$ as
$t\to\infty$. We choose a convenient model form for $J(t)$ to be
\begin{equation}
J(t)=\frac{J_0}{e^{kt}+1}.
\end{equation}
so that $\frac{d}{dt}J(t)=-J_0 k e^{kt}/(e^{kt}+1)^2$.


\begin{figure}[H]
\includegraphics[width=9cm]{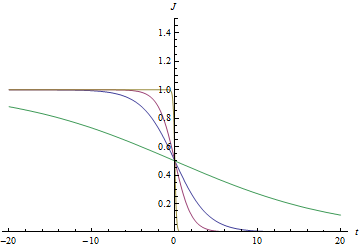}
\caption{\label{fig:jt1}Figure of the coupling $J(t)$ over time. These
lines represent $k=0.1, 0.5, 1, 10$ respectively.}
\end{figure}


We see that when $k\to 0$, $\frac{d}{dt}J(t)$ will go to zero; when
$k\to \infty$, it will become a Heaviside step function with a jump at
$t=0$ (see Fig.\ref{fig:jt1} ). These 2 limits describe two
complementary processes. The $k=0$ limit simply adiabatically decouple
the system from the bath, whereas the $k\rightarrow \infty$ limit
suddenly switches off the coupling.  For this sudden decoupling case,
the result should be the same as what we get from previous sections
since the central system stops evolving after the turn-off. But for
the adiabatic decoupling case, things are more complicated - we need
to solve the problem explicitly - we will do this in the long-time
limit.

\subsection{Time-dependent solutions}

With the time-dependent $J(t)$ we can follow the same
Jordan-Wigner/Bogoliubov procedure as before, and now transform the
Hamiltonian to
\begin{equation}
H=\sum_n
2E_n(\gamma_n^{\dag}\gamma_n-\frac{1}{2})+\sum\frac{iJ(t)}{\sqrt{N}}(\gamma_n^{\dag}-\gamma_n),
\end{equation}
where $E_n=V$ for model 1, and $E_n=V\cos k_n$ for model 2. Then we
have
\begin{equation}
H_n^{\pm}=\begin{pmatrix}
2E_n & \mp\frac{i J(t)}{\sqrt{N}}\\
\pm\frac{i J(t)}{\sqrt{N}} & 0
\end{pmatrix}.
\end{equation}

Our goal is to solve this Hamiltonian in each
$\{\gamma_n^{\dag}|0\rangle,|0\rangle\}$ subspace. A vector
$(a(t),b(t))$ in this subspace evolves as
\begin{equation}
-i\begin{pmatrix}
\frac{d }{d t}a(t)\\
\frac{d }{d t}b(t)
\end{pmatrix}=\begin{pmatrix}
2E_n & \mp\frac{i J(t)}{\sqrt{N}}\\
\pm\frac{i J(t)}{\sqrt{N}} & 0
\end{pmatrix}\begin{pmatrix}
a(t)\\
b(t)
\end{pmatrix}.
\end{equation}
so that
\begin{eqnarray}
&a(t)=\mp\frac{\sqrt{N}}{J(t)} \dot{b}(t),\\
&\ddot{b(t)}-(iE_n+\frac{\dot{J}(t)}{J(t)})\dot{b}(t)+\frac{J^2(t)}{N}b(t)=0\label{eq:RDE1}.
\end{eqnarray}

After some derivation (see Appendix), we get a result for $\kappa(t)$
from this. The long time asymptotic limit $t\rightarrow \infty$ of
$\kappa(t)$ is then:
\begin{equation}
\kappa(t)=\prod_n (1-\frac{2 J^2}{N E_n^2}).
\end{equation}

For the two models we then have:

\vspace{2mm}

(i) {\it Model 1}: Here we have
$H_{bath}=V\sum_{i=0}^{N-1}\sigma_i^z\sigma_{i+1}^z$, and $E_n=V$ for
all subspaces. Then the decoherence factor is
\begin{equation}\label{eq:kappa3}
\kappa_1(t)=(1-\frac{2 J^2}{N E_n^2})^N=e^{-2J_0^2}.
\end{equation}

\vspace{2mm}

(ii) {\it Model 2}: Now we have
$H_{bath}=V\sum_{i=0}^{N-1}(\sigma_i^{+}\sigma_{i+1}^{-}+h.c.)$, and
$E_n=V\cos k_n$ for each subspaces. The decoherence factor is
\begin{equation}
\kappa_2(t)=e^{-\frac{1}{2\pi}\int_{0}^{2\pi}dk\frac{2J^2}{V^2\cos^2
k}}\to 0.
\end{equation}

\vspace{2mm}

Thus, in this adiabatically decoupled limit, we find ``partial
decoherence" for model 1 ($\kappa\rightarrow e^{-2J^2}$)and ``complete
decoherence"  for model 2 ($\kappa\rightarrow 0$). Apparently, system
1 ends up in a mixed state and the coherence is partially lost.

However, appearances are deceptive in this case. To see this, we can
imagine reversing the coupling back to its original value. We
therefore consider  what happens for a general slow-varying $J(t)$.
In this case \eqref{eq:RDE1} still holds, and we can consider both
models 1 and 2.

\vspace{2mm}

(i){\it Model 1}: Here again we have $E_n=V$. If the coupling is
varying slowly, $\dot{J(t)}\ll J(t)E_n$, we can safely neglect the
$dJ/dt$ term in \eqref{eq:RDE1} and we have
\begin{widetext}
\begin{equation}
\begin{split}
b(t)=&C_1e^{-iVt/2-iVt/2\sqrt{1+\frac{4J^2(t)}{NV^2}}}+C_2e^{-iVt/2+iVt/2\sqrt{1+\frac{4J^2(t)}{NV^2}}},\\
a(t)=&-\frac{i2C_1
\sqrt{N}}{J(t)}(V+V\sqrt{1+\frac{4J^2(t)}{NV^2}})e^{-iVt/2-iVt/2\sqrt{1+\frac{4J^2(t)}{NV^2}}}\\
&-\frac{i2C_2
\sqrt{N}}{J(t)}(V-V\sqrt{1+\frac{4J^2(t)}{NV^2}})e^{-iVt/2+iVt/2\sqrt{1+\frac{4J^2(t)}{NV^2}}}.
\end{split}
\end{equation}
\end{widetext}

We can see that in this adiabatic limit, the result has no dependence
on $dJ/dt$. The decoherence factor $\kappa_1(t)$ should therefore be
solely dependent on the form of $J(t)$. This means that if one slowly
changes the coupling strength back to its original value,
$\kappa_1(t)$ would go back to its original value regardless of the
apparent loss of decoherence we obtained in \eqref{eq:kappa3}. One can
thus recover the full coherence by decoupling the model 1 to its
environment and then recoupling it back adiabatically. The coherence
is not truly lost into the environment, but simply encoded there in
the entanglement between the central spin and the first environmental
spin. It can be transferred back to the central spin simply by
reversing the interaction.

\vspace{2mm}

(ii){\it Model 2}:  In model 2 we cannot neglect the $\dot{J}(t)$
term, since $E_n=V\cos k_n$ can be zero. The slow varying condition
$dJ/dt \ll J(t)E_n$ breaks down for the modes in the center of the
band.  Actually it is easy to prove that $\kappa(t)$ must stay at $0$
after adiabatically decoupling from the environment, and that never
come back to its initial state even if one reverses the interaction.
In this case, the coherence is truly lost through the modes near
$k_n\approx\frac{\pi}{2}, \frac{3\pi}{2}$, once we take the limit $N
\rightarrow \infty$.

This result is of course typical of a system coupled to an oscillator
bath. One can easily generalize the above considerations to a
finite-temperature bath.


\section{Discussion}


It is always helpful, in discussing some general physical phenomenon,
to have simple toy models which illustrate basic features of the
physics. Decoherence is of course very general - it has been studied
intensively in contexts ranging from black hole physics
\cite{deco-BHole} and infrared features of quantum field theory
\cite{IR-QFT}, as well as in a huge variety of condensed matter
systems. Understanding  decoherence is also key to doing quantum
computation.

As noted in the introduction, there exist thorough studies of
decoherence in several well-known models \cite{AJL87,weiss,PS00}. Even
in a simple central spin model, there are several decoherence
mechanisms in play \cite{PS00}; these mechanisms generalize to more
general central systems like particles hopping on rings \cite{ZZ10} or
moving on hyperlattices \cite{PS06}. In all these models one finds no
connection between decoherence and dissipation, quite unlike the case
for oscillator bath models like the spin-boson model
\cite{AJL87,weiss}.

The two models here, taken in conjunction with results for the central
spin model \cite{PS00}, make it very clear why this is the case. Let
us recall the form of the central spin model Hamiltonian: one has
 \begin{equation}
 \label{eq:H} H
\;=\; H_{CS} + H_{SB},
 \end{equation}
where the central spin part is
 \begin{eqnarray}
H_{CS}& =& [\Delta_{o} \hat{\tau}_x  e^{i\sum_k (\mbox{\boldmath
$\phi$}_k + \mbox{\boldmath
$\alpha$}_{k}\cdot \mbox{\boldmath $\sigma$}_k)} + H.c.]\\
&& \qquad\qquad +
 (\epsilon_o + \sum_k \mbox{\boldmath $\lambda$}^{k} \cdot
\mbox{\boldmath $\sigma$}_k) \hat{\tau}_z
\end{eqnarray}
and the spin bath Hamiltonian is
 \begin{equation}
H_{SB} \;=\; \sum_{k}{\bf h}_k \cdot \mbox{\boldmath $\sigma$}_k
\;+\; \sum_{k,k'} V_{kk'}^{\alpha \beta} \sigma_{k}^{\alpha}
\sigma_{k'}^{\beta}.
 \end{equation}
in which $\sigma_k$ is again the bath spin operator.

Some of the terms in this Hamiltonian are of no special interest here.
The Berry phase coupling $\mbox{\boldmath $\phi$}_k$ and the
topological phase $\mbox{\boldmath
$\alpha$}_{k}$ (which leads to topological decoherence \cite{PS00})
naturally arise in the truncation of the central system to a 2-level
system, but are usually quite small in realistic systems.

The `orthogonality blocking' vector coupling $\mbox{\boldmath
$\lambda$}^{k}$ is not so easily dismissed - it leads to `precessional
decoherence', which is often the major source of decoherence for
systems at low $T$. Nevertheless we shall drop it, noting that further
work, in which it is included, will be of considerable interest when
it comes to applications of these toy models.

Finally, we note that the interaction $V_{kk'}^{\alpha \beta}$ between
bath spins is often very small (particularly if the spins are nuclear,
or if the bath spins are widely separated in space). On longer time
scales it does play an essential role (in creating 'fluctuational
decoherence' \cite{PS00}), but at shorter timescales it can be treated
as negligible.

Let us therefore assume that we now drop all these interactions, so
that $V_{kk'}^{\alpha \beta} = 0$, and  that $\mbox{\boldmath
$\phi$}_k = \mbox{\boldmath$\alpha$}_{k} = 0$. Then this reduced
Central (CS) spin model is simply a generalization of our `model 1',
in which the central spin now couples simultaneously to all of the
bath spins instead of just the first one in the chain. It is known
from the solution to this reduced CS model \cite{PS00} that
decoherence exists in it, provided one averages over the bath spin
states. However, if we do not make such an average, then an initially
pure state gradually distributes itself over the $N$ spins in the bath
\cite{coxThesis}. The time evolution will be periodic, just as for
`model 1', but now the period is exponentially long (with timescale
$\propto \exp N^{1/3}$); after this timescale, the information `comes
back' to reconstitute the original central spin state.

Thus model 1 is just the maximum simplification that one can make of
the original CS model. Because model 1 is essentially just a 2-spin
problem, we get the simple results given above. If we generalize model
1 to allow coupling of the central qubit to some reduced set of $M$
bath spins, then the basic result of the calculation will now be clear
- quantum correlations and entanglement will simply be shared over
time between the central qubit and the $M$ bath spins, with
entanglement between different bath spins mediated by the central
spin. Very accurate solutions for a problem of this kind (which is
very relevant to experiment) can be found using entanglement
correlator methods \cite{cox18,coxThesis}.

Model 2 is not a simplification of any central spin model; instead, as
already noted, it is just a special case of the spin-boson model. If
we had wanted to, we could have simply derived the form of the
Caldeira-Leggett \cite{AJL87} spectral function $J(\omega)$ for this
system, and proceeded from there. Note however that if the number $N$
of spins in the chain is finite, we will also get, in model 2, the
quantum analogue of Poincar$\acute{e}$ recurrences - but in the
calculations given above, we have let $N \rightarrow \infty$. In model
2, the spin wave modes with $k_n\approx \frac{\pi}{2}, \frac{3\pi}{2}$
serve the purpose of carrying quantum information away from the
central spin.

Obviously there is a lot more one can say about these 2 simple models.
It will be of interest, for example, to analyse the higher-spin
`entanglement correlators' \cite{cox18,coxThesis} in model 2, and to
see what happens as one adds fields $\{ {\bf h}_k \}$ acting on the
bath spins, and/or gradually `switches on' the interaction between the
bath spins. Many models are then possible, all of them generalizations
of the 2 simple central spin models discussed here.

Finally, we note that one can also generalize these 2 models to deal
with a central system in which a particle hops around some site
Hamiltonian. This is nothing but a 1-dimensional `polaron' model -
with a spin bath background, it is a type of spin polaron system. It
is also related to the original model discussed by Feynman for quantum
computation \cite{RPF-QIP}. The central system now has a Hamiltonian
 \begin{equation}
 \label{eq:H_line}
H_o \;=\;- \sum_{<ij>} \left[ t_{ij} c_i^{\dagger} c_j \: e^{i
A_{ij}^{0}}  + H.c.\right] + \sum_j \varepsilon_j c_j^{\dagger} c_j
 \end{equation}
in which we also admit a Berry phase $A_{ij}^{0}$ connected to hopping
between nearest-neighbour sites.

The most general coupling to the spin bath then takes the form
\begin{eqnarray}
 V_{int} \;=\; &\sum_{k}^{N}&
[\; \sum_j \mbox{\boldmath $F$}_j^k(\mbox{\boldmath $\sigma$}_k)
\hat{c}_{j}^{\dagger}\hat{c}_j  \nonumber\\
 \;\;\;\;\;\;\;\;&+& \sum_{ij} (
\mbox{\boldmath $G$}_{ij}^k(\mbox{\boldmath $\sigma$}_k)
\hat{c}_{i}^{\dagger}\hat{c}_j +H.c.) ]
 \label{spin-z}
\end{eqnarray}
with both diagonal coupling $\mbox{\boldmath $F$}_j^k$
and non-diagonal coupling $\mbox{\boldmath $G$}_{ij}^k$ of the hoping
particle to the $k$-th bath spin.

We see immediately that if we choose a simple site-diagonal coupling
$\mbox{\boldmath $F$}^k(\mbox{\boldmath $\sigma$}_k)$ on each site,
making all these couplings the same for each site, and also let the
site energy $\varepsilon_j$ be the same for each site, then we have
another toy model for which concrete calculations can be done. A
related model was treated some time ago \cite{ZZ10}, in which the line
was closed to form a ring.

This work was supported by the National Science and Engineering
Council of Canada. We thank Tim Cox for discussions.

\vspace{3mm}

{\bf DEDICATION}: This paper is dedicated to my long-time colleague
Gordon Semenoff, whom I first met while I was a postdoc, and he
already a professor; at that time I was rather awed by him! Since then
he has been a welcome source of advice, wisdom, and friendship, with
similar views on many things; and we even once wrote a paper
\cite{GWS06} together! I wish him many more years happily carving his
own path, in the inimitable way he has done up to now.

\appendix


\section{Appendix}\label{ap:spin chain}


Here we derive two key results used in the text, viz., the effect of
the non-commutativity of the $\{ H_n \}$ in the discussion of model 1,
and the derivation of the solution used for adiabatic switching for
both models.

\subsection{Neglecting Commutators}

In \eqref{eq:commutator}, we noticed that the $H_n'$s do not commute
with each other; in fact
\begin{equation}
[H_n,H_{n'}]=-\frac{2J^2}{N}(\gamma_{k_n}^{\dag}-\gamma_{k_n})(\gamma_{k_{n'}}^{\dag}-\gamma_{k_{n'}}).
\end{equation}
We can  use the Zassenhaus formula \cite{CPA:CPA3160070404}  to get
\begin{equation}\begin{split}
e^{i\sum_n H_n t}&=\prod_n e^{i H_n t} \prod_{n_1> n_2}
e^{\frac{t^2}{2}[H_{n_1},H_{n_2}]}\\
&\times \prod_{n_1\geq n_2\geq n_3}e^{\frac{-i
t^3}{6}([H_{n_1},[H_{n_2},H_{n_3}]]+[[H_{n_1},H_{n_2}],H_{n_3}])}....\end{split}
\end{equation}
Although the Hamiltonian is not block diagonal in each
$\{\gamma_{k_n}^{\dag}|0\rangle,|0\rangle\}$ subspace, if we restrict
our model to the weak interaction region, i.e. $J\ll 1$, we can omit
this commutator since it always comes into the final expression in
higher order. For example , for the
$\prod_{n_1, n_2} e^{\frac{t^2}{2}[H_{n_1},H_{n_2}]}$ term, we could
expand it in powers of $1/N$ and take expectation values
\begin{equation}
\begin{split}
&\mbox{tr}_b\left(\prod_{n_1> n_2}
e^{\frac{t^2}{2}[H_{n_1}^{+},H_{n_2}^{+}]}\prod_{n_1, n_2}
e^{\frac{t^2}{2}[H_{n_1}^{-},H_{n_2}^{-}]}\right)\\
&=\mbox{tr}_b
\left(e^{-\frac{J^2t^2}{N}\sum_{n_1>n_2}(\gamma_{k_{n_1}}^{\dag}-\gamma_{k_{n_1}})(\gamma_{k_{n_2}}^{\dag}-\gamma_{k_{n_2}})}\right)\\
&\approx 1-\frac{J^4t^4}{N^2} N(N-1).
\end{split}
\end{equation}
As we can see in the main text this already happens in the $J^2$
order. Therefore in the weakly interacting region, our calculation can
be a good approximation to the exact result.

\subsection{Adiabatic Decoupling}

The goal is to solve the asymptotic solution to the following equation
\begin{eqnarray}
&a(t)=\mp\frac{\sqrt{N}}{J(t)} \dot{b}(t)\\
&\ddot{b(t)}-(iE_n+\frac{\dot{J}(t)}{J(t)})\dot{b}(t)+\frac{J^2(t)}{N}b(t)=0.
\end{eqnarray}

We first study the evolution of the state $(0,1)$. Actually, the state
$(1,0)$ can be treated in the same way just with an opposite sign of
$J(t)$ and $E_n$. Then we have the initial condition
\begin{eqnarray}
b(t_0)&=&1\\
\dot{b}(t_0)&=&0.
\end{eqnarray}
Define the following quantity for simplification, keeping in mind that
$\frac{J_0^2}{N}$ is very small
\begin{equation}
\epsilon=\frac{J_0^2}{N}.
\end{equation}

The general solution of the equation \eqref{eq:RDE1} is
\begin{widetext}
\begin{equation}
b(t)=C_1e^{-iE_nt/2-iE_n\sqrt{1+4\epsilon/E_n^2}/2}(1+e^{kt})^{i\sqrt{\epsilon}/k}f_1(-e^{kt})
\;\;+\;\;
C_2e^{-iE_nt/2+iE_n\sqrt{1+4\epsilon/E_n^2}/2}(1+e^{kt})^{i\sqrt{\epsilon}/k}f_2(-e^{kt})
\end{equation}

Here we have
\begin{eqnarray}
f_1(x)&=&_2F_1
\left(1+\frac{iE_n}{2k}+\frac{i\sqrt{\epsilon}}{k}-\frac{iE_n}{2k}\sqrt{1+\frac{4\epsilon}{E_n^2}};
\;
-\frac{iE_n}{2k}+\frac{i\sqrt{\epsilon}}{k}-\frac{iE_n}{2k}\sqrt{1+\frac{4\epsilon}{E_n^2}};
\;1-\frac{iE_n}{2k}\sqrt{1+\frac{4\epsilon}{E_n^2}}; \; x \right)\\
f_2(x)&=&_2F_1
\left(1+\frac{iE_n}{2k}+\frac{i\sqrt{\epsilon}}{k}+\frac{iE_n}{2k}\sqrt{1+\frac{4\epsilon}{E_n^2}};
\;
-\frac{iE_n}{2k}+\frac{i\sqrt{\epsilon}}{k}+\frac{iE_n}{2k}\sqrt{1+\frac{4\epsilon}{E_n^2}};
\; 1+\frac{iE_n}{2k}\sqrt{1+\frac{4\epsilon}{E_n^2}}; \;x \right)
\end{eqnarray}

We set our initial time $t_0\to -\infty$, then $e^{kt}\to 0$ and
\begin{equation}
f_{1,2}(-e^{kt})\to 1.
\end{equation}

Therefore
\begin{equation}
\begin{split}
b(t_0)&\;\;=\;\; 1\\
&\;\;=\;\; C_1e^{-iE_nt_0/2-iE_nt_0\sqrt{1+4\epsilon/E_n^2}/2}\;+\;
C_2e^{-iE_nt_0/2+iE_nt_0\sqrt{1+4\epsilon/E_n^2}/2};\end{split}
\end{equation}
\begin{equation}\begin{split}
\dot{b}(t_0)\;\;=\;\;&1\\
\;\;=\;\;&C_1(-iE_n/2-iE_n\sqrt{1+4\epsilon/E^2}/2)e^{-iE_nt_0/2-iE_nt_0\sqrt{1+4\epsilon/E_n^2}/2}\\
&+C_2(-iE_n/2+iE_n\sqrt{1+4\epsilon/E_n^2}/2)e^{-iE_nt_0/2+iE_nt_0\sqrt{1+4\epsilon/E_n^2}/2}
\end{split}
\end{equation}

Then we can get the constant $C_1, C_2$ as
\begin{eqnarray}
C_1&=&\frac{1}{2}e^{iE_nt_0/2+iE_nt_0\sqrt{1+4\epsilon/E_n^2}/2}(-1+(1+\frac{4\epsilon}{E_n^2})^{-\frac{1}{2}})\\
C_2&=&\frac{1}{2}e^{iE_nt_0/2-iE_nt_0\sqrt{1+4\epsilon/E_n^2}/2}(1+(1+\frac{4\epsilon}{E_n^2})^{-\frac{1}{2}}).
\end{eqnarray}

After we get the solution , we can study its behavior when
$t\to\infty$. We use the fact that when $t\to\infty$
\begin{eqnarray}
f_1(-e^{kt})&\to & Ae^{iE_nt/2-i\sqrt{\epsilon}
t+\frac{itE_n}{2}\sqrt{1+4\epsilon/E_n^2}/2}\\
f_2(-e^{kt})&\to & Be^{iE_nt/2-i\sqrt{\epsilon}
t-\frac{itE_n}{2}\sqrt{1+4\epsilon/E_n^2}/2}
\end{eqnarray}
with
\begin{eqnarray}
A&=&\frac{\Gamma(1-\frac{iE_n}{k}\sqrt{1+4\epsilon/E_n^2})\Gamma(1+\frac{iE_n}{k})}{\Gamma(1+\frac{iE_n}{2k}-\frac{i\sqrt{\epsilon}}{k}-\frac{iE_n}{k}\sqrt{1+4\epsilon/E_n^2})}
\times
\frac{1}{\Gamma(1+\frac{iE_n}{2k}+\frac{i\sqrt{\epsilon}}{k}-\frac{iE_n}{k}\sqrt{1+4\epsilon/E_n^2})}\\
B&=&\frac{\Gamma(1+\frac{iE_n}{k}\sqrt{1+4\epsilon/E_n^2})\Gamma(1+\frac{iE_n}{k})}{\Gamma(1+\frac{iE_n}{2k}-\frac{i\sqrt{\epsilon}}{k}+\frac{iE_n}{k}\sqrt{1+4\epsilon/E_n^2})}\times
\frac{1}{\Gamma(1+\frac{iE_n}{2k}+\frac{i\sqrt{\epsilon}}{k}+\frac{iE_n}{k}\sqrt{1+4\epsilon/E_n^2})}.
\end{eqnarray}
\end{widetext}

Recalling that $\epsilon=\frac{J_0^2}{N}$ is a very small number, we
can expand the solution according in powers of  $\epsilon$. The
results are as follows:\\\\

\paragraph*{Zeroth order}: if $\epsilon=0$, then $C_1\to 0$, $C_2\to
1$, and
\begin{equation}
B\to
\frac{\gamma(1+\frac{iE_n}{k})\gamma(1+\frac{iE_n}{k})}{\gamma(1+\frac{iE_n}{k})\gamma(1+\frac{iE_n}{k})}\to
1.
\end{equation}
Thus we find that $b(t)\to 1$, which is exactly what we expected. \\\\
\paragraph*{First order}
Since $\sqrt{1+\frac{4\epsilon}{E_n^2}}\approx
1+\frac{2\epsilon}{E^2}$, we have
\begin{eqnarray}
C_1 &\to & e^{iE_n t_0}(1+\frac{i\epsilon
t}{E_n})(-\frac{\epsilon}{E^2})=-\frac{\epsilon}{E_n^2}e^{iE_n t_0}\\
C_2 &\to & e^{-i\epsilon
t_0/E_n^2}(1-\frac{\epsilon}{E_n})=1-\frac{\epsilon}{E_n^2}-\frac{i\epsilon
t}{E_n^2}.
\end{eqnarray}

As a result
\begin{widetext}
\begin{eqnarray}
A &\;\;\to\;\; &\Gamma(1-\frac{iE_n}{k})\Gamma(1+\frac{iE_n}{k})\times
\Bigg[1
-\frac{i\epsilon}{kE_n}\left(\frac{\Gamma'(1-iE_n/k)}{\Gamma(1-iE_n/k)}+i\Gamma'(1)\right)\nonumber
\;+\; \frac{\epsilon}{E_n^2}(\Gamma'(1)-\Gamma''^2(1))\Big]\\
B &\;\;\to\;\; &
1-\frac{\Gamma'^2(1+\frac{iE_n}{k})}{\Gamma(1+\frac{iE_n}{k})^2}\frac{\epsilon}{k^2}+\frac{\Gamma''(1+\frac{iE}{k})}{\Gamma(1+i\frac{E_n}{k})}\frac{\epsilon}{k^2};
\end{eqnarray}
so that for $b(t)$ we get the behaviour
\begin{equation}
b(t)\;\;\to\;\;  1-\frac{\epsilon}{k^2}e^{iE_n
t_0}\Gamma(1-\frac{iE_n}{k})\Gamma(1+\frac{iE_n}{k})-\frac{\epsilon}{E_n^2}-\frac{i\epsilon
t}{E_n^2}
-\frac{\epsilon}{k^2}\frac{\Gamma'^2(1+\frac{iE_n}{k})}{\Gamma(1+\frac{iE_n}{k})^2}+\frac{\epsilon}{k^2}\frac{\Gamma''(1+\frac{iE}{k})}{\Gamma(1+i\frac{E_n}{k})}+O(\epsilon^{\frac{3}{2}})
\end{equation}
\end{widetext}

Then the trace of $\mbox{tr}(e^{iH_{+}t}e^{iH_{-}t})$ is
$|b(t)|^2-|a(t)|^2$. Since $|b(t)|^2-|a(t)|^2=1$,
$\mbox{tr}(e^{iH_{+}t}e^{iH_{-}t})=2|b(t)|^2-1$.
For the adiabatic decoupling we mentioned in the main text, we take
$k\to 0$, then $\frac{d}{dt}J(t)\to 0$. The system is slowly decoupled
from the bath. We use the fact that when $|z|\to\infty$
\begin{eqnarray}
(\ln \Gamma(z))'&=&\ln z+\frac{z-1/2}{z}+O(z^{-2})\\
(\ln
\Gamma(z))''&=&\frac{\Gamma''(z)}{\Gamma(z)}-\left(\frac{\Gamma'(z)}{\Gamma(z)}\right)^2\\
(\ln \Gamma(z))''&=&\frac{1}{z}+\frac{1}{2z^2}+O(z^{-3})\\
\end{eqnarray}

We can get the factor $b(t)$ as well as the decoherence factor
$\kappa_n$ for this subspace; one gets
\begin{equation}
 |b(t)|^2\to \; 2(1-\frac{\epsilon}{E_n^2})
\end{equation}
and also that
\begin{equation}
\kappa_n(t)\;=\; \mbox{tr}(e^{iH_n^{+}t}e^{iH_n^{-}t})\;\;\to\;\;
1-\frac{2 J_0^2}{N E_n^2}
\end{equation}

This is the result used in the main text to get $\kappa(t)$ for models
1 and 2.


\end{document}